\documentclass[conference]{IEEEtran}
\usepackage{subfigure}
\usepackage{amsmath}
\usepackage{graphicx,amssymb,lineno,bm,subfigure}
\usepackage{algorithm}
\usepackage{algorithmic}

\newtheorem{thm}{\textbf{Theorem}}
\newtheorem{rmk}{\textbf{Remark}}

\newtheorem{proof}{\textbf{Proof}}
\newtheorem{lma}{\textbf{Lemma}}
\IEEEoverridecommandlockouts
\begin{document}

%
\title{\huge{Power-Delay Tradeoff in Multi-User Mobile-Edge Computing Systems}}
\author{\IEEEauthorblockN{Yuyi Mao{$^\dagger$}, Jun Zhang{$^\dagger$}, S.H. Song{$^\dagger$}, and K. B. Letaief{$^{\dagger\ast}$}, \emph{Fellow, IEEE}}
\IEEEauthorblockA{$^\dagger$Dept. of ECE, The Hong Kong University of Science and Technology, Clear Water Bay, Hong Kong\\
$^{\ast}$Hamad bin Khalifa University, Doha, Qatar\\
Email: \{ymaoac, eejzhang, eeshsong, eekhaled\}@ust.hk}
\thanks{This work is supported by the Hong Kong Research Grants Council under Grant No. 16200214.}
}

\maketitle

\begin{abstract}
Mobile-edge computing (MEC) has recently emerged as a promising paradigm to liberate mobile devices from increasingly intensive computation workloads, as well as to improve the quality of computation experience. In this paper, we investigate the tradeoff between two critical but conflicting objectives in multi-user MEC systems, namely, the power consumption of mobile devices and the execution delay of computation tasks. A power consumption minimization problem with task buffer stability constraints is formulated to investigate the tradeoff, and an online algorithm that decides the local execution and computation offloading policy is developed based on Lyapunov optimization. Specifically, at each time slot, the optimal frequencies of the local CPUs are obtained in closed forms, while the optimal transmit power and bandwidth allocation for computation offloading are determined with the Gauss-Seidel method. Performance analysis is conducted for the proposed algorithm, which indicates that the power consumption and execution delay obeys an $\left[O\left(1\slash V\right),O\left(V\right)\right]$ tradeoff with $V$ as a control parameter. Simulation results are provided to validate the theoretical analysis and demonstrate the impacts of various parameters to the system performance.
\end{abstract}

%
\IEEEpeerreviewmaketitle
\section{Introduction}
The increasing popularity of smart mobile devices is driving the development of mobile applications, which can be computation-intensive, e.g., interactive online gaming, face recognition and 3D modeling. This poses more stringent requirements on the quality of computation experience, which cannot be easily satisfied by the limited processing capability of mobile devices. As a result, new solutions to handle the explosive computation demands and the ever-increasing computation quality requirements are emerging \cite{Gubbi1309}. Mobile-edge computing (MEC) is such a promising technique to release the tension between the computation-intensive applications and the resource-limited mobile devices \cite{ETSI14}. Different from conventional cloud computing systems, where remote public clouds are utilized, MEC offers computation capability within the radio access network. Therefore, by offloading the computation tasks from the mobile devices to the MEC servers, the quality of computation experience, including the device energy consumption and execution latency, can be greatly improved \cite{Satyanarayanan0910}.

Nevertheless, the efficiency of computation offloading highly depends on the wireless channel conditions, as offloading tasks requires effective data transmission. Therefore, computation offloading policies for MEC systems have attracted significant attention in recent years \cite{WZhang1309}-\cite{Kwak1512}. For applications with strict deadline requirements, the local execution energy consumption was minimized by adopting dynamic voltage and frequency scaling (DVFS) techniques, and the energy consumption for computation offloading was optimized using data transmission scheduling in \cite{WZhang1309}. In \cite{Munoz1510}, joint allocation of communication and computational
resources for femto-cloud computing systems was proposed, where each computation task should be completed before its deadline. In \cite{YMao1612}, a dynamic computation offloading policy was developed for MEC systems with energy harvesting devices under a strict execution delay requirement. Besides, a decentralized computation offloading algorithm was proposed to minimize the computation overhead for multi-user MEC systems in \cite{XChen1504}.

Imposing strict execution delay constraints makes the computation offloading design more tractable, as only short-term performance, e.g., the performance for executing a single task, needs to be considered. However, it may be impractical for applications that can tolerate a certain period of execution latency, such as multi-media streaming. For such type of applications, the long-term system performance is more relevant, where the coupling among the randomly arrived tasks cannot be ignored. In order to minimize the long-term average energy consumption, a stochastic control algorithm was proposed in \cite{DHuang1206}, which determines the offloaded software components. In \cite{JLiu1607}, a delay-optimal stochastic task scheduling algorithm was developed for single-user MEC systems. Moreover, an online task scheduling algorithm was proposed to investigate the energy-delay tradeoff for MEC systems with a multi-core mobile device in \cite{ZJiang1512}, and this study was later extended to scenarios with heterogeneous types of mobile applications in \cite{Kwak1512}. Unfortunately, existing works only focused on single-user MEC systems, and the design methodologies for multi-user MEC systems remain unknown.

In this paper, we consider a general MEC system with multiple mobile devices, where computation tasks arrive at the mobile devices in a stochastic manner. Joint design of local execution and computation offloading strategies will be investigated. With multiple devices, the design becomes much more challenging as intelligent management of the radio resources for computation offloading, e.g., the transmit power and available spectrum, is needed. We formulate a power consumption minimization problem with task buffer stability constraints. An online algorithm is proposed based on Lyapunov optimization, which decides the CPU-cycle frequencies for local execution, and the transmit power and bandwidth allocation for computation offloading. In particular, the optimal CPU-cycle frequencies are obtained in closed forms, while the optimal transmit power and bandwidth allocation are determined by the Gauss-Seidel method. Performance analysis is conducted for the proposed algorithm, which explicitly characterizes the tradeoff between the power consumption of the mobile devices and the execution delay. Simulation results verify the theoretical analysis and demonstrate that the proposed algorithm is capable of controlling the power consumption and execution delay performance in multi-user MEC systems.

The organization of this paper is as follows. We introduce the system model in Section II. The power consumption minimization problem is formulated in Section III, and an online local execution and computation offloading policy is developed in Section IV. Simulation results will be shown in Section V, and we will conclude this paper in Section VI.

\section{System Model}
\vspace{-8pt}
\begin{figure}[h]
\centering
\includegraphics[width=0.42\textwidth]{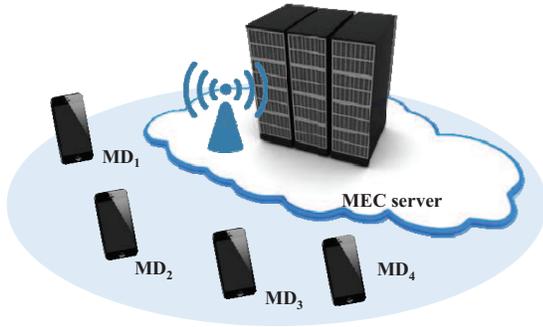}
\vspace{-15pt}
\caption{A mobile-edge computing system with four mobile devices (MDs).}
\label{sysmodelMEC}
\end{figure}

We consider a mobile-edge computing (MEC) system as shown in Fig. \ref{sysmodelMEC}, where $N$ mobile devices running computation-intensive applications are assisted by an MEC server. The MEC server could be a small data center installed at a wireless access point deployed by the telecom operator. Therefore, it can be accessed by the mobile devices through wireless channels, and will execute the computation tasks on behalf of the mobile devices \cite{Satyanarayanan0910,WZhang1309}. By offloading part of the computation tasks to the MEC server, the mobile devices could enjoy a higher level of quality of computation experience \cite{Satyanarayanan0910}.

The available system bandwidth is $w$ Hz, which is shared by the mobile devices, and the noise power spectral density at the receiver of the MEC server is denoted as $N_{0}$. Time is slotted and the time slot length is $\tau$. For convenience, we denote the index sets of the mobile devices and the time slots as $\mathcal{N}\triangleq\{1,\cdots,N\}$ and $\mathcal{T}\triangleq\{0,1,\cdots\}$, respectively.

\subsection{Computation Task and Task Queueing Models}
We assume the mobile devices are running fine-grained tasks \cite{Kwak1512}: At the beginning of the $t$th time slot, $A_{i}\left(t\right)$ (bits) of computation tasks arrive at the $i$th mobile device, which can be processed starting from the $\left(t+1\right)$th time slot. Without loss of generality, we assume the $A_{i}\left(t\right)$'s in different time slots are independent and identically distributed (i.i.d.) within $\left[A_{i,\min},A_{i,\max}\right]$ with $\mathbb{E}\left[A_{i}\left(t\right)\right]=\lambda_{i},i\in\mathcal{N}$.

In each time slot, part of the computation tasks of the $i$th mobile device, denoted as $D_{l,i}\left(t\right)$, will be executed at the local CPU, while $D_{r,i}\left(t\right)$ bits of the computation tasks will be offloaded to and executed by the MEC server. The arrived but not yet executed tasks will be queued in the task buffer at each mobile device, and the queue lengths of the task buffers at the beginning of the $t$th time slot are denoted as $\mathbf{Q}\left(t\right)\triangleq\left[Q_{1}\left(t\right),\cdots,Q_{N}\left(t\right)\right]$ with $\mathbf{Q}\left(0\right)=\mathbf{0}$, where $Q_{i}\left(t\right)$ evolves according to the following equation:
\begin{equation}
Q_{i}\left(t+1\right)=\max\{Q_{i}\left(t\right)-D_{\Sigma,i}\left(t\right),0\}+A_{i}\left(t\right),t\in \mathcal{T}.
\label{bufferdynamics}
\end{equation}
In (\ref{bufferdynamics}), $D_{\Sigma,i}\left(t\right)=D_{l,i}\left(t\right)+D_{r,i}\left(t\right)$ is the amount of tasks departing from the task buffer at the $i$th device in time slot $t$.

\subsection{Local Execution Model}
In order to process one bit of task input at the $i$th mobile device, $L_{i}$ CPU cycles will be needed, which depends on the types of applications and can be obtained by off-line measurements \cite{Miettinen10}. Denote the scheduled CPU-cycle frequency for the $i$th mobile device in the $t$th time slot as $f_{i}\left(t\right)$, which cannot exceed $f_{i,\max}$. Thus, $D_{l,i}\left(t\right)$ can be expressed as
\begin{equation}
D_{l,i}\left(t\right)=\tau f_{i}\left(t\right)L_{i}^{-1}.
\label{Dlocal}
\end{equation}
Accordingly, the power consumption for local execution at the $i$th mobile device is given by
\begin{equation}
p_{l,i}\left(t\right)=\kappa f^{3}_{i}\left(t\right),
\label{Plocal}
\end{equation}
where $\kappa$ is the effective switched capacitance related to the chip architecture \cite{Burd9608}.

\subsection{MEC Server Execution Model}
To offload the computation tasks for MEC server execution, the input bits of the tasks need to be delivered to the MEC server. For simplicity, we assume the MEC server is equipped with an $N$-core high-speed CPU so that it can execute $N$ different applications in parallel, and the processing latency at the MEC server is negligible. We leave the investigation of more general MEC servers to our future work.

The wireless channels between the mobile devices and the MEC server are i.i.d. frequency-flat block fading. Denote the small-scale fading channel power gain from the $i$th mobile device to the MEC server at the $t$th time slot as $h_{i}\left(t\right)$, which is assumed to have a finite mean value, i.e., $\mathbb{E}\left[h_{i}\left(t\right)\right]=\overline{h_{i}}<\infty$. Thus, the channel power gain from the $i$th mobile device to the MEC server can be represented by $H_{i}\left(t\right)=h_{i}\left(t\right)g_{0}\left(d_{0}\slash d_{i}\right)^{\theta}$, where $g_{0}$ is the path-loss constant, $d_{0}$ is the reference distance, $\theta$ is the path-loss exponent, and $d_{i}$ is the distance from mobile device $i$ to the MEC server. Hence, the amount of offloaded tasks at the $i$th mobile device in time slot $t$ is given by
\begin{equation}
D_{r,i}\left(t\right)=
\begin{cases}
\alpha_{i}\left(t\right)w\tau\log_{2}\left(1+\frac{H_{i}\left(t\right)p_{{\rm{tx}},i}\left(t\right)}{\alpha_{i}\left(t\right)N_{0}w}\right), &\alpha_{i}\left(t\right)>0\\
0, &\alpha_{i}\left(t\right)=0,
\end{cases}
\label{Dremote}
\end{equation}
where $p_{{\rm{tx}},i}\left(t\right)$ is the transmit power with the maximum value of $p_{i,\max}$, and $\alpha_{i}\left(t\right)$ is the portion of bandwidth allocated to the $i$th mobile device. Denote $\bm{\alpha}\left(t\right)\triangleq\left[\alpha_{1}\left(t\right),\cdots,\alpha_{N}\left(t\right)\right]$ as the bandwidth allocation vector, which should be chosen from the feasible set $\mathcal{A}$ \cite{ZWang1510}, i.e.,
\begin{equation}
\bm{\alpha}\left(t\right)\in\mathcal{A}\triangleq \bigg\{\bm{\alpha}\in\mathbb{R}_{+}^{N}\big|\sum\nolimits_{i\in\mathcal{N}}\alpha_{i}\leq 1\bigg\}.
\label{alphadomain}
\end{equation}

\section{Problem Formulation}
In this section, we will first introduce the performance metrics, namely, the power consumption of the mobile devices and the average queue lengths of the task buffers. A power consumption minimization problem will then be formulated to facilitate the investigation of the power-delay tradeoff.

The average power consumption of the mobile devices, including the power consumed by the local CPUs and the transmit power for computation offloading, can be expressed as
\begin{equation}
\overline{P}=\lim_{T\rightarrow \infty} \frac{1}{T}\mathbb{E}\left[\sum_{t=0}^{T-1}P\left(t\right)\right],
\label{netwpwr}
\end{equation}
where $P\left(t\right)\triangleq \sum_{i\in\mathcal{N}}\left(p_{{\rm{tx}},i}\left(t\right)+p_{l,i}\left(t\right)\right)$.

According to the \emph{Little's Law} \cite{Queuetheory}, the execution delay is proportional to the average queue length of the task buffer. Hence, we adopt the average queue length of the task buffer as a measurement of the execution delay, which can be written as
\begin{equation}
\overline{Q}_{i}=\lim_{T\rightarrow \infty}\frac{1}{T}\mathbb{E}\left[\sum_{t=0}^{T-1}Q_{i}\left(t\right)\right],i\in\mathcal{N}.
\end{equation}

Denote $\mathbf{f}\left(t\right)\triangleq\left[f_{1}\left(t\right),\cdots,f_{N}\left(t\right)\right]$ and $\mathbf{p}_{{\rm{tx}}}\left(t\right)\triangleq \left[p_{{\rm{tx}},1}\left(t\right),\cdots,p_{{\rm{tx}},N}\left(t\right)\right]$. Thus, the power consumption minimization problem is formulated as
\begin{align}
&\mathbf{P}_{1}: \min_{\mathbf{f}\left(t\right),\mathbf{p}_{\rm{tx}}\left(t\right),\bm{\alpha}\left(t\right)} \overline{P}\nonumber\\
&\ \ \ \ \ \ \ \ \ \ \ \ \mathrm{s.t.\ \ \ }\bm{\alpha}\left(t\right)\in\mathcal{A},t\in\mathcal{T}\label{alphadomain2}\\
&\ \ \ \ \ \ \ \ \ \ \ \ \ \ \ \ \ \ \ 0\leq f_{i}\left(t\right)\leq f_{i,\max},i\in\mathcal{N},t\in\mathcal{T} \label{freqconstraint}\\
&\ \ \ \ \ \ \ \ \ \ \ \ \ \ \ \ \ \ \ 0\leq p_{{\rm{tx}},i}\left(t\right)\leq p_{i,\max},i\in\mathcal{N},t\in\mathcal{T} \label{txconstraint}\\
&\ \ \ \ \ \ \ \ \ \ \ \ \ \ \ \ \ \ \lim_{t\rightarrow \infty}\frac{\mathbb{E}\left[|Q_{i}\left(t\right)|\right]}{t}=0,i\in\mathcal{N}, \label{stabilityconstraint}
\end{align}
where (\ref{freqconstraint}) and (\ref{txconstraint}) are the CPU-cycle frequency constraint and the transmit power constraint, respectively. (\ref{stabilityconstraint}) requires the task buffers to be mean rate stable \cite{Neely10}, which ensures that all the arrived computation tasks can be executed with finite delay. In general, $\mathbf{P}_{1}$ is a stochastic optimization problem, for which, the CPU-cycle frequency, the transmit power as well as the bandwidth allocation need to be determined for each device at each time slot. This problem is difficult to solve as the optimal decisions are temporally correlated. Also, a joint consideration on the local execution and computation offloading strategies is needed, as both of them affect the system performance. Besides, the spatial coupling of the bandwidth allocation among different mobile devices poses another challenge.

Instead of solving $\mathbf{P}_{1}$ directly, we consider $\mathbf{P}_{2}$, which is a modified version of $\mathbf{P}_{1}$ by replacing set $\mathcal{A}$ in (\ref{alphadomain}) by set $\tilde{\mathcal{A}}$, with $\tilde{\mathcal{A}}$ defined as
\begin{equation}
\tilde{\mathcal{A}}=\bigg\{\bm{\alpha}\in\mathbb{R}_{+}^{N}|\sum\nolimits_{i\in\mathcal{N}}\alpha_{i}\leq 1,\alpha_{i}\geq\epsilon_{A},i\in\mathcal{N}\bigg\}.
\end{equation}
With such modification, the departure function of MEC server execution, $D_{r,i}\left(t\right)$, is continuous and differentiable with respect to $\bm{\alpha}\left(t\right)\in\tilde{\mathcal{A}}$. In addition, the optimal value of $\mathbf{P}_{2}$ is larger but can be made arbitrarily close to that of $\mathbf{P}_{1}$ by setting $\epsilon_{A}$ ($\epsilon_{A}\in\left(0,1\slash N\right)$) to be sufficiently small. Furthermore, any feasible solution for $\mathbf{P}_{2}$ is also feasible for $\mathbf{P}_{1}$. Thus, we will focus on $\mathbf{P}_{2}$ in the remainder of this paper.

\section{Online Local Execution and Computation Offloading Policy}
In this section, we will propose an online local execution and computation offloading policy to solve $\mathbf{P}_{2}$ based on Lyapunov optimization \cite{Neely10}, where a deterministic problem needs to be solved at each time slot. We will then analyze the performance of the proposed algorithm and reveal the power-delay tradeoff in multi-user MEC systems.

\subsection{Lyapunov Optimization-Based Online Algorithm}
To present the algorithm, we first define the Lyapunov function as
\begin{equation}
L\left(\mathbf{Q}\left(t\right)\right)=\frac{1}{2}\sum_{i\in\mathcal{N}}Q_{i}^{2}\left(t\right).
\label{Lyvfunc}
\end{equation}
Thus, the Lyapunov drift function can be written as
\begin{equation}
\Delta\left(\mathbf{Q}\left(t\right)\right)=
\mathbb{E}\left[L\left(\mathbf{Q}\left(t+1\right)\right)-L\left(\mathbf{Q}\left(t\right)\right)|\mathbf{Q}\left(t\right)\right].
\label{Lyvdrift}
\end{equation}
Accordingly, the Lyapunov drift-plus-penalty function can be expressed as
\begin{equation}
\Delta_{V}\left(\mathbf{Q}\left(t\right)\right)
=\Delta\left(\mathbf{Q}\left(t\right)\right)+V\cdot\mathbb{E}\left[P\left(t\right)|\mathbf{Q}\left(t\right)\right],
\label{Lyvdriftpenalty}
\end{equation}
where $V$ (${\rm{bits}}^{2}\cdot {\rm{W}}^{-1}$) is a control parameter in the proposed algorithm. We find an upper bound of $\Delta_{V}\left(\mathbf{Q}\left(t\right)\right)$ under any feasible $\mathbf{f}\left(t\right)$, $\mathbf{p}_{\rm{tx}}\left(t\right)$, and $\bm{\alpha}\left(t\right)$, as specified in Lemma \ref{Lyvdriftpenaltyboundlma}.

\begin{lma}
For arbitrary $\mathbf{f}\left(t\right),\mathbf{p}_{\rm{tx}}\left(t\right),\bm{\alpha}\left(t\right)$ such that $\forall i\in\mathcal{N}$, $f_{i}\left(t\right)\in\left[0,f_{i,\max}\right]$, $p_{{\rm{tx}},i}\left(t\right)\in\left[0,p_{i,\max}\right]$, and $\bm{\alpha}\left(t\right)\in\tilde{\mathcal{A}}$, $\Delta_{V}\left(t\right)$ is upper bounded by\
\begin{equation}
\begin{split}
\Delta_{V}\left(\mathbf{Q}\left(t\right)\right)&\leq - \mathbb{E}\left[\sum_{i\in\mathcal{N}}Q_{i}\left(t\right)\left(D_{\Sigma,i}\left(t\right)-A_{i}\left(t\right)\right)|\mathbf{Q}\left(t\right)\right]\\
&+V\cdot \mathbb{E}\left[P\left(t\right)|\mathbf{Q}\left(t\right)\right]+C,
\end{split}
\label{Lyvdriftpenaltybound}
\end{equation}
where $C$ is a constant.
\label{Lyvdriftpenaltyboundlma}
\end{lma}
\begin{proof}
Proof is omitted due to space limitation.
\end{proof}

The main idea of the proposed online local execution and computation offloading policy is to minimize the upper bound of $\Delta_{V}\left(\mathbf{Q}\left(t\right)\right)$ in the right-hand side of (\ref{Lyvdriftpenaltybound}) greedily at each time slot. By doing so, the amount of tasks waiting in the task buffers can be maintained at a small level. Meanwhile, the power consumption of the mobile devices can be minimized. The proposed algorithm is summarized in Algorithm \ref{Algframework}, where a deterministic optimization problem $\mathbf{P}_{\rm{PTS}}$ needs to be solved at each time slot. It is worthy to note that the objective function of $\mathbf{P}_{\rm{PTS}}$ corresponds to the right-hand side of (\ref{Lyvdriftpenaltybound}), and all the constraints in $\mathbf{P}_{2}$ except the stability constraints in (\ref{stabilityconstraint}) are retained in $\mathbf{P}_{\rm{PTS}}$. The optimal solution for $\mathbf{P}_{\rm{PTS}}$ will be developed in the next subsection.

\begin{algorithm}[h]
\caption{Lyapunov Optimization-Based Online Local Execution and Computation Offloading Policy}
\label{alg1}
\begin{algorithmic}[1]
\STATE At the beginning of the $t$th time slot, obtain $\{Q_{i}\left(t\right)\}$, $\{H_{i}\left(t\right)\}$, and $\{A_{i}\left(t\right)\}$.
\STATE Determine $\mathbf{f}\left(t\right), \mathbf{p}_{\rm{tx}}\left(t\right)$ and $\bm{\alpha}\left(t\right)$ by solving
\begin{align}
&\mathbf{P}_{\rm{PTS}}:\min_{\mathbf{f}\left(t\right),\mathbf{p}_{\rm{tx}}\left(t\right),\bm{\alpha}\left(t\right)} -\sum_{i\in\mathcal{N}}Q_{i}\left(t\right)D_{\Sigma,i}\left(t\right) +V\cdot P\left(t\right)\nonumber \\
&\ \ \ \ \ \ \ \ \ \ \ \ \ \ \ \mathrm{s.t.}\ \ \ \ \bm{\alpha}\left(t\right)\in\tilde{\mathcal{A}}, (\ref{freqconstraint})\ {\rm{and}}\ (\ref{txconstraint}).\nonumber
\nonumber
\end{align}
\STATE Update $\{Q_{i}\left(t\right)\}$ according to (\ref{bufferdynamics}) and set $t=t+1$.
\end{algorithmic}
\label{Algframework}
\end{algorithm}

\subsection{Optimal Solution For $\mathbf{P}_{\rm{PTS}}$}

In this subsection, we will derive the optimal CPU-cycle frequencies, transmit powers and bandwidth allocation vector for $\mathbf{P}_{\rm{PTS}}$.

\textbf{Optimal CPU-cycle Frequencies:} It is straightforward to show that the optimal CPU-cycle frequency for the $i$th mobile device in time slot $t$ can be obtained by solving
\begin{equation}
\mathbf{SP}_{1}:\min_{0\leq f_{i}\left(t\right)\leq f_{i,\max}} -Q_{i}\left(t\right) \tau f_{i}\left(t\right) L_{i}^{-1} + V \kappa f^{3}_{i}\left(t\right),
\end{equation}
and its optimal solution is achieved at either the stationary point of the objective function or one of the boundary points, which is given by
\begin{equation}
f_{i}^{\star}\left(t\right)=\min\bigg\{f_{i,\max},\sqrt{\frac{Q_{i}\left(t\right)\tau }{3\kappa V L_{i}}}\bigg\},i\in\mathcal{N}.
\end{equation}
\begin{rmk}
Note that $f_{i}^{\star}\left(t\right)$ increases with $Q_{i}\left(t\right)$ as it is desirable to execute more tasks in order to keep the queue length of the task buffer small. Besides, $f^{\star}_{i}\left(t\right)$ decreases with both $V$ and $L_{i}$: With a larger value of $V$, the weight of the power consumption becomes larger, and thus the local CPU slows down its frequency to reduce power consumption; with a larger value of $L_{i}$, local execution becomes less efficient as more CPU cycles are needed to process per bit of task input, which leads to a smaller CPU-cycle frequency.
\end{rmk}

\textbf{Optimal Transmit Power and Bandwidth Allocation:} After decoupling $\mathbf{f}\left(t\right)$ from $\mathbf{P}_{\rm{PTS}}$, the optimal $\mathbf{p}^{\star}_{\rm{tx}}\left(t\right)$ and $\bm{\alpha}^{\star}\left(t\right)$ can be obtained by solving
\begin{equation}
\begin{split}
&\mathbf{SP}_{2}:\min_{\bm{\alpha}\left(t\right),\mathbf{p}_{{\rm{tx}}}\left(t\right)}-\sum_{i\in\mathcal{N}}Q_{i}\left(t\right)D_{r,i}\left(t\right)
+V\sum_{i\in\mathcal{N}}p_{{\rm{tx}},i}\left(t\right)\\
&\ \ \ \ \ \ \ \ \ \ \ {\mathrm{s.t.}}\ \ \ \ 0\leq p_{{\rm{tx}},i}\left(t\right)\leq p_{i,\max},i\in\mathcal{N}\\
&\ \ \ \ \ \ \ \ \ \ \ \ \ \ \ \ \ \ \ \bm{\alpha}\left(t\right)\in\tilde{\mathcal{A}}.
\end{split}
\end{equation}
It is not difficult to identify that $\mathbf{SP}_{2}$ is a convex optimization problem. However, generic convex algorithms suffer from relatively high complexity as they are developed for general convex problems and do not make use of the problem structures \cite{Boyd04}. Motivated by this, we propose to solve $\mathbf{SP}_{2}$ by optimizing the transmit power and bandwidth allocation in an alternating manner, where in each iteration, the optimal transmit powers are obtained in closed forms and the optimal bandwidth allocation is determined by the \emph{Lagrangian method}. Since $\mathbf{SP}_{2}$ is jointly convex with respect to $\mathbf{p}_{{\rm{tx}}}\left(t\right)$ and $\bm{\alpha}\left(t\right)$, and its feasible region is a Cartesian product of those of $\mathbf{p}_{\rm{tx}}\left(t\right)$ and $\bm{\alpha}\left(t\right)$, the alternating minimization procedure is guaranteed to converge to the global optimal solution, which is termed as the \emph{Gauss-Seidel method} in literature \cite{Grippo0004}.

\textbf{1) Optimal Transmit Power:} For a fixed bandwidth allocation vector $\bm{\alpha}\left(t\right)$, the optimal transmit power for the $i$th mobile device can be obtained by solving
\begin{equation}
\mathbf{P}_{\rm{PWR}}:\min_{0\leq p_{{\rm{tx}},i}\left(t\right)\leq p_{i,\max}}-Q_{i}\left(t\right)D_{r,i}\left(t\right)+Vp_{{\rm{tx}},i}\left(t\right),
\end{equation}
whose optimal solution is achieved at either the stationary point of the objective function or one of the boundary points similar to $\mathbf{SP}_{1}$, and it is given in closed form by
\begin{equation}
\begin{split}
&p^{\star}_{{\rm{tx}},i}\left(t\right)=\\
&\min\bigg\{\alpha_{i}\left(t\right)w\max\bigg\{\frac{Q_{i}\left(t\right)\tau}{\ln2 \cdot V}-\frac{N_{0}}{H_{i}\left(t\right)},0\bigg\},p_{i,\max}\bigg\},i\in\mathcal{N}.
\end{split}
\label{optPWR}
\end{equation}

\textbf{2) Optimal Bandwidth Allocation:} For a fixed transmit power vector $\mathbf{p}_{{\rm{tx}}}\left(t\right)$, the optimal bandwidth allocation can be obtained by solving the following problem:
\begin{align}
&\mathbf{P}_{\rm{BW}}:\min_{\bm{\alpha}\left(t\right)\in\tilde{\mathcal{A}}} -\sum_{i\in\mathcal{N}}Q_{i}\left(t\right)D_{r,i}\left(t\right),
\end{align}
which is more challenging as the bandwidth allocation decision is coupled among different mobile devices. Fortunately, the Lagrangian method offers an effective solution for $\mathbf{P}_{\rm{BW}}$. Specifically, the partial Lagrangian can be written as
\begin{equation}
\mathcal{L}\left(\bm{\alpha}\left(t\right),\lambda\right)=-\sum_{i\in\mathcal{N}}Q_{i}\left(t\right)D_{r,i}\left(t\right)+\lambda \left(\sum_{i\in\mathcal{N}}\alpha_{i}\left(t\right)-1\right),
\end{equation}
where $\lambda\geq 0$ is the Lagrangian multiplier associated with constraint $\sum_{i\in\mathcal{N}}\alpha_{i}\left(t\right)\leq 1$. Based on the Karush-Kuhn-Tucker (KKT) conditions, the optimal bandwidth allocation $\bm{\alpha}^{\star}\left(t\right)$ and the optimal Lagrangian multiplier $\lambda^{\star}$ should satisfy the following equation set:
\begin{equation}
\begin{cases}
&\alpha_{i}^{\star}\left(t\right)=\max\{\epsilon_{A},\mathcal{R}_{i}\left(\lambda^{\star}\right)\},i\in\mathcal{N},\lambda^{\star}>0\\
&\sum_{i\in\mathcal{N}}\alpha_{i}^{\star}\left(t\right)=1.
\end{cases}
\label{KKT}
\end{equation}
In (\ref{KKT}), if $p_{{\rm{tx}},i}^{\star}\left(t\right)=0$, $\mathcal{R}_{i}\left(\lambda\right)\triangleq \epsilon_{A}$; otherwise, $\mathcal{R}_{i}\left(\lambda\right)$ denotes the root of $Q_{i}\left(t\right)\frac{d D_{r,i}\left(t\right)}{d \alpha_{i}\left(t\right)}=\lambda$ for $\lambda>0$, which is positive and unique as $\frac{dD_{r,i}\left(t\right)}{d\alpha_{i}\left(t\right)}$ decreases with $\alpha_{i}\left(t\right)$. Thus, it suggests a bisection search over $\left[\lambda_{L},\lambda_{U}\right]$ for the optimal $\lambda^{\star}$, where $\lambda_{L}=\max_{i\in\mathcal{N}}Q_{i}\left(t\right)\frac{dD_{r,i}\left(t\right)}{d\alpha_{i}\left(t\right)}|_{\alpha_{i}\left(t\right)=1}$, and $\lambda_{U}$ satisfies $\sum_{i\in\mathcal{N}}\max\{\epsilon_{A},\mathcal{R}_{i}\left(\lambda_{U}\right)\}<1$. Hence, $\mathcal{R}_{i}\left(\lambda\right)$ can be obtained by a bisection search over $\left(0,1\right]$, and the searching process for the optimal $\lambda^{\star}$ will be terminated when $|\sum_{i\in\mathcal{N}}\max\{\epsilon_{A},\mathcal{R}_{i}\left(\lambda\right)\}-1|<\xi$, where $\xi$ is the accuracy of the algorithm. Details of the Lagrangian method for $\mathbf{P}_{\rm{BW}}$ are summarized in Algorithm \ref{Lagrangianmtd}.

\begin{algorithm}[h] 
\caption{Lagrangian Method for $\mathbf{P}_{\rm{BW}}$} 
\begin{algorithmic}[1] 
\STATE Set $\xi = 10^{-7}$, $\lambda_{U}=\lambda_{L}$, $l=0$, $I_{\max}=200$, $\beta=1.5$, $\epsilon_{A}=10^{-4}$.
\STATE Set $\alpha_{i}\left(t\right)=\max\{\epsilon_{A},\mathcal{R}_{i}\left(\lambda_U\right)\},i\in\mathcal{N}$.
\STATE \textbf{While} {$\sum_{i\in\mathcal{N}}\alpha_{i}\left(t\right)\geq 1$} \textbf{do}
\STATE \hspace{10pt} $\lambda_{U}=\beta\cdot \lambda_{U}$.
\STATE \hspace{10pt} Set $\alpha_{i}\left(t\right)=\max\{\epsilon_{A},\mathcal{R}_{i}\left(\lambda_U\right)\},i\in\mathcal{N}$.
\STATE \textbf{Endwhile}
\STATE \textbf{While} {$|\sum_{i\in\mathcal{N}}\alpha_{i}\left(t\right)-1|\geq   \xi$} and $l\leq I_{\max}$ \textbf{do}
\STATE \hspace{10pt} $\tilde{\lambda}=\frac{1}{2}\left(\lambda_{L}+\lambda_{U}\right)$ and $l=l+1$.
\STATE \hspace{10pt} Set $\alpha_{i}\left(t\right)=\max\{\epsilon_{A},\mathcal{R}_{i}\left(\tilde{\lambda}\right)\},i\in\mathcal{N}$.
\STATE \hspace{10pt} \textbf{If} $\sum_{i\in\mathcal{N}}\alpha_{i}\left(t\right)>1$ \textbf{then}
\STATE \hspace{20pt} $\lambda_{L}=\tilde{\lambda}$.
\STATE \hspace{10pt} \textbf{Else}
\STATE \hspace{20pt} $\lambda_{U}=\tilde{\lambda}$.
\STATE \hspace{10pt} \textbf{Endif}
\STATE \textbf{Endwhile}
\end{algorithmic}
\label{Lagrangianmtd}
\end{algorithm}

\begin{rmk}
One main benefit of the proposed online algorithm is that it does not require prior information on the computation task arrival and wireless channel fading processes, which makes it also applicable for unpredictable environments. Besides, the proposed algorithm is of low complexity, as at each time slot, the optimal CPU-cycle frequencies are obtained in closed forms, while the computation offloading policy is determined by an efficient alternating minimization algorithm. Furthermore, as will be shown in the next subsection, the achievable performance of the proposed algorithm can be analytically characterized and thus facilitates the analysis on the power-delay tradeoff in multi-user MEC systems.
\end{rmk}

\subsection{Performance Analysis}
In this subsection, we will provide the main theoretical result in this paper, which characterizes the upper bounds for the power consumption of the mobile devices and the average sum queue length of the task buffers. Also, the tradeoff between the power consumption and execution delay will be revealed.

\begin{thm}
Assume that $\mathbf{P}_{2}$ is feasible, we have:
\begin{itemize}
\item The average power consumption of the mobile devices under the proposed algorithm satisfies:
\begin{equation}
\overline{P}\leq P^{\rm{opt}}_{\Sigma}+C\cdot V^{-1},
\end{equation}
where $P_{\Sigma}^{\rm{opt}}$ is the optimal value of $\mathbf{P}_{2}$.
\item For arbitrary $i\in\mathcal{N}$, $Q_{i}\left(t\right)$ is mean rate stable.
\item Suppose there exist $\epsilon>0$ and $\Psi\left(\epsilon\right)$ ($\Psi\left(\epsilon\right)>P_{\Sigma}^{\rm{opt}}$) that satisfy the Slater conditions \cite{Neely10}, then the average sum queue lengths of the task buffers satisfies:
  \begin{equation}
    \sum\nolimits_{i\in\mathcal{N}}\overline{Q}_{i}\leq \left[C+V\left(\Psi\left(\epsilon\right)-P_{\Sigma}^{\rm{opt}}\right)\right]\cdot{\epsilon^{-1}}.
   \end{equation}
\end{itemize}
\label{performanalysis}
\end{thm}
\begin{proof}
Proof is omitted due to space limitation.
\end{proof}

\begin{rmk}
Theorem \ref{performanalysis} shows that under the proposed online local execution and computation offloading policy, the worst-case power consumption of the mobile devices decreases inversely proportional to $V$, while the upper bound of the execution delay increases linearly with $V$, i.e., there exists an $\left[O\left(1\slash V\right),O\left(V\right)\right]$ tradeoff between these two objectives. Thus, we can balance the power consumption and execution delay by adjusting $V$: For delay-sensitive types of applications, we can use a small value of $V$; while for energy-sensitive networks and delay-tolerant applications, a large value of $V$ can be adopted.
\end{rmk}

\section{Simulation Results}
In simulations, we assume $N$ mobile devices are located at an equal distance of $150$ m from the MEC server. The small-scale fading channel power gains are exponentially distributed with unit mean. Besides, $\kappa = 10^{-27}$, $\tau=1$ ms, $w=10$ MHz, $N_{0}=-174$ dBm\slash Hz, $g_{0}=-40$ dB, $d_{0}=1$ m, $\theta=4$, $f_{i,\max}=1$ GHz, $p_{i,\max}=500$ mW, $A_{i}\left(t\right)$ is uniformly distributed within $\left[0,A_{i,\max}\right]$, and $L_{i}=737.5$ cycles\slash bit, $i\in\mathcal{N}$ \cite{Miettinen10}. The simulation results are averaged over 5000 time slots.

\vspace{-5pt}
\begin{figure}[h]
\centering
\includegraphics[width=0.42\textwidth]{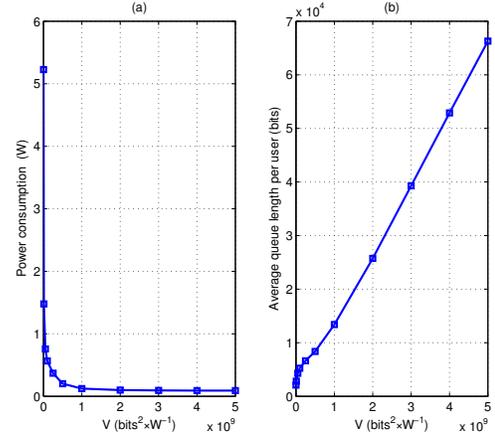}
\vspace{-10pt}
\caption{Power consumption of the mobile devices/average queue length per user vs. the control parameter $V$, $N=5$ and $A_{i,\max}=4$ kbits.}
\label{FIGPWRQV}
\end{figure}

We first show the relationship between the power consumption of the mobile devices/average queue length of the task buffers and the control parameter $V$ in Fig. \ref{FIGPWRQV}. We see from Fig. \ref{FIGPWRQV}a) that the power consumption decreases inversely proportional to $V$ and converges to $P_{\Sigma}^{\rm{opt}}$ when $V$ is sufficiently large. Meanwhile, as shown in Fig. \ref{FIGPWRQV}b), the average queue length of the task buffers increases linearly with $V$ and becomes unbounded when $V$ goes to infinity. These results verify the $\left[O\left(1\slash V\right),O\left(V\right)\right]$ tradeoff between the power consumption and execution delay as shown in Theorem \ref{performanalysis}.

\begin{figure}[!t]
\centering
\includegraphics[width=0.42\textwidth]{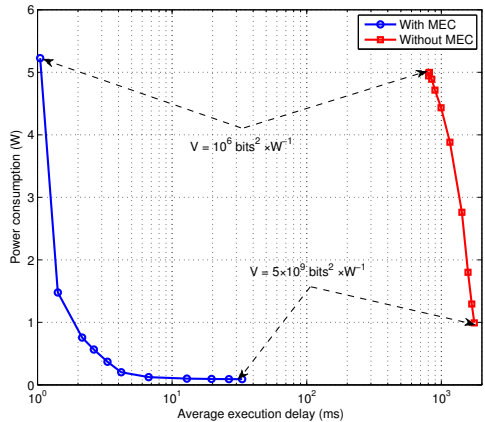}
\vspace{-10pt}
\caption{Power consumption of the mobile devices vs. execution delay for systems with and without MEC, $N=5$ and $A_{i,\max}=4$ kbits.}
\label{FIGWITHWITHOUTMEC}
\end{figure}

In Fig. \ref{FIGWITHWITHOUTMEC}, we show the relationship between the power consumption and execution delay for scenarios with and without MEC\footnote{The average execution delay is calculated by $\sum_{i\in\mathcal{N}}\overline{Q}_{i}\slash \sum_{i\in\mathcal{N}}\lambda_{i}$ (time slots) according to the Little's Law.}. It is observed that by increasing $V$ from $10^{6}$ to $5\times10^{9}\ {\rm{bit}}^{2}\cdot {\rm{W}}^{-1}$, the power consumption of the mobile devices decreases significantly for both cases. However, the behaviors of the execution delay are substantially different: With MEC, the execution delay decreases sharply from 33.2 to 1.05 ms as $V$ decreases, while without MEC, the execution delay has minor changes at around $10^{3}$ ms. This is because without the aid of the MEC server, the devices cannot stabilize their task buffers even with a small $V$, where the local CPUs operate at their maximum frequencies. Therefore, we verify the benefits of MEC for improving the quality of computation experience.

\begin{figure}[!t]
\centering
\includegraphics[width=0.42\textwidth]{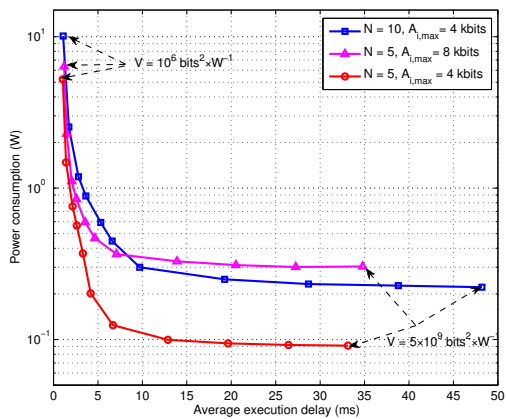}
\vspace{-10pt}
\caption{Power consumption of the mobile devices vs. execution delay.}
\label{FIGTraoff}
\end{figure}

By varying $A_{i,\max}$ and $N$, we show the relationship between the power consumption and execution delay in Fig. \ref{FIGTraoff}. In general, the average execution delay increases as the power consumption decreases, which indicates that a proper $V$ should be chosen to balance the two desirable objectives. For instance, with $N=5$ and $A_{i,\max}=4$ kbits, if the average execution delay requirement is 20 ms, $V=3\times 10^{9}$ ${\rm{bits}}^{2}\cdot {\rm{W}}^{-1}$ can be chosen, and the power consumption will be 0.1 W. Besides, with a given execution delay, the power consumption increases with the computation task arrival rate (the number of mobile devices), which agrees with the intuitions as the workload of the MEC system becomes heavier, more power is needed to stabilize the task buffers. In addition, when $V$ goes to infinity, doubling the computation task arrival rates results in a higher power consumption than doubling the number of mobile devices, which is due to the increased multi-user diversity gain and the availability of extra local CPUs.

\section{Conclusions}
In this paper, we investigated the power-delay tradeoff in a multi-user mobile-edge computing system. A power consumption minimization problem with task buffer stability constraints was formulated, and an online algorithm that decides the local execution and computation offloading policy was derived based on Lyapunov optimization. Performance analysis was conducted for the proposed algorithm, which explicitly characterizes the $\left[O\left(1\slash V\right),O\left(V\right)\right]$ tradeoff between the power consumption and execution delay performance. Simulation results validated the theoretical analysis, and showed that the proposed algorithm is capable of balancing the power consumption of the mobile devices and the quality of computation experience. For future investigation, it would be interesting to extend the findings in this work to scenarios with fairness considerations among multiple devices.

\end{document}